\documentclass[aps,pra,showpacs,twoside,10pt,floatfix,nofootinbib,longbibliography,twocolumn]{revtex4-1}
\usepackage[colorlinks=true, citecolor=red, urlcolor=blue ]{hyperref}
\usepackage{epsfig,newlfont,amssymb,amsfonts,amsmath,bm,subfigure,palatino,mathtools,amsthm,braket,soul,enumitem,color,times,comment,geometry,xcolor}
\usepackage[normalem]{ulem}
\definecolor{boldcolor}{HTML}{7541C0}
\definecolor{boldcolor1}{HTML}{4A2166}
\definecolor{boldcolor2}{HTML}{6C3391}
\definecolor{boldcolor3}{HTML}{366E8A}
\definecolor{boldcolor4}{HTML}{69B2D6}

\newcommand{\PROP}[1]{\vspace{0.1cm}\noindent\textbf{\textcolor{boldcolor1}{$\blacksquare$ Proposition #1.}}}

\begin{document}

\title{Bridging the daemonic gap en route to charge multi-mode batteries via a single auxiliary}
\author{Chandrima B. Pushpan, Amit Kumar Pal}
\affiliation{Department of Physics, Indian Institute of Technology Palakkad, Palakkad 678 623, India}
\date{\today}

\begin{abstract}
We show that harnessing daemonic advantage is possible while charging a quantum battery by first time-evolving the battery collectively with an auxiliary charger, followed by an energy extraction via tracing out the charger. We define the difference between the minimum daemonic ergotropy and the maximum ergotropy of the battery as the daemonic gap at the time where the ergotropy of the battery is maximum. Considering a harmonic mode as the battery and a qubit as the auxiliary charger interacting  via Jaynes-Cummings interaction,  we show that the daemonic gap can be closed for specific initial passive states of the battery, including the ground state, truncated mixtures of low-lying states, and canonical thermal states. We further define the daemonic band as the difference between the maximum and the minimum daemonic ergotropy, and show that starting from the ground state of the single-mode battery, the daemonic band collapses. We also show that  achieving a complete charging, along with accessing the full daemonic band, is possible for the single-mode battery using all of the other initial states by repeating the charging cycle and maximizing ergotropy in every round.  We extend the battery-charger design to a multi-mode resonator battery and a qudit charger, and demonstrate, for a double-mode battery and a qutrit charger, that repeating the charging cycle is vital for simultaneously charging individual modes of the collective battery.   
\end{abstract}

\maketitle

\section{Introduction}
\label{sec:intro}

Quantum batteries~\cite{Alicki2013,Campaioli2018,Campaioli2024} have recently attracted immense attention, exploring a variety of possible designs and working principles~\cite{Santos2019,Crescente2020,Mitchison2021,Crescente2022,Zhu2023,Downing2023,Ahmadi2024,Song2024} probing optimal performance~\cite{Julia-Farre2020,Gyhm2022}, the speed and efficiency of charging and discharging~\cite{Crescente2020,Ge2023,Lai2024,Shastri2025,Evangelakos2025}, the possibility of extracting quantum advantage~\cite{Andolina2025}, as well as the role of quantum correlations in the performance of the battery~\cite{Hovhannisyan2013,Binder2015,Campaioli2017,Sen2021,Shi2022,Gyhm2024}. It has emerged as a topic of interest in broader scientific domains, being influenced by other fields such as artificial intelligence~\cite{Rosa2023}, many-body physics~\cite{Le2018,Ferraro2018,Andolina2019,Rossini2019,Rossini2020,Ghosh2020,Ghosh2021,Ghosh2022,Arjmandi2022,Konar2022_a,Arjmandi2023,Mitra2024,Puri2024arX,Perciavalle2025,Lu2025}, quantum chaos~\cite{Romero2025}, non-Hermitian physics~\cite{Konar2024}, continuous variable systems~\cite{Konar2024_a}, material science~\cite{Camposeo2025}, and quantum computation~\cite{Kurman2025}. The possibility of implementing an energy storage device in systems compatible with quantum technology has motivated realization of quantum batteries using trapped ions~\cite{Zhang2025}, nuclear magnetic resonance systems~\cite{Joshi2022},  quantum dots~\cite{Wenniger2023}, transmons~\cite{Hu2022,Dou2023}, nitrogen vacancy center~\cite{Niu2024}, and organic semiconductors~\cite{Quach2022}.

In a bipartite setup involving a battery $b$ and an auxiliary charger $a$, a suitable initial state $\rho_{ab}^{\text{in}}=\rho_a^{\text{in}}\otimes\rho_b^{\text{in}}$ of the battery-charger duo, constituted as the product of a \emph{passive state}~\cite{Alicki2013,Campaioli2018,Campaioli2024,Pusz1978,Singh2021,Allahverdyan2004} $\rho_b^{\text{in}}$ of the battery, and an educated choice of the auxiliary charger state $\rho_a^{\text{in}}$, is first taken through a time-evolution $U_{ab}$ generated by a joint battery-charger Hamiltonian, resulting in a charging of the battery. This is followed by an extraction of the battery energy via suitable operations on the auxiliary charger at a judiciously chosen time $t$, where the time-evolution and the work extraction together is referred to as a \emph{charging cycle}. As explored in a couple of recent works~\cite{Yan2023,Zhang2024}, the energy-extraction is possible by a projection measurement on the auxiliary charger. Specifically, a particular measurement outcome $k$, occurring with the probability $p^k$, is chosen post-measurement, with the corresponding post-measured battery state $\rho_b^k$ having non-zero extractable work $\mathcal{E}^k$ as quantified by the ergotropy~\cite{Allahverdyan2004}, leading to a probabilistic work extraction protocol. This allows for the possibility of sequential application of the charging cycle to enhance charging, as the battery is naturally initiated in the state $\rho_b^k$ for the upcoming charging cycle. The strategy for repeating the charging cycle is applied in~\cite{Yan2023} to charge a quantum battery in the form of a single-mode resonator  using an auxiliary qubit charger.

Stopping the time evolution at a specific time $t$ and using a measurement for work extraction, one can also exploit the daemonic advantage due to the information gained from the measurement by computing the  ergotropy averaged over the post-measured ensemble on the battery corresponding to a specific measurement, denoted by $\overline{\mathcal{E}}=\sum_kp^k\mathcal{E}^k$ and referred to as the daemonic ergotropy~\cite{Francica2017}. This has been explored  by a series of recent works~\cite{Francica2017,Bernards2019,Morrone2023,Barra2024,Kua2025} (cf.~\cite{Chaki2023_arx,Chaki2024_arx,Chaki2024_arx_a}), and the idea of daemonic gain has been introduced~\cite{Francica2017} quantifying the daemonic advantage due to measuring the charger over discarding the same, maximized over all possible measurements on the charger. However, translating the sequential application of the charging cycle to the daemonic sector demands careful consideration, as a natural choice for the battery initial state prior to each charging cycle is lacking.

Extraction of energy is also possible via discarding the auxiliary charger altogether by integrating out all its degrees of freedom, which is equivalent to mixing $\rho_b^k$ with probability $p^k$ corresponding to an arbitrary projection measurement on the auxiliary charger, leaving the battery in a state $\rho_b=\sum_kp^k\rho_b^k$. While the ergotropy, $\mathcal{E}$, of $\rho_b$ quantifies the maximum work attainable using this protocol, and one can, in principle, repeat the charging cycle starting from $\rho_b$ as the initial state of the battery, the daemonic advantage is lost, as $\overline{\mathcal{E}}\geq\mathcal{E}$ for all local projection measurements on the auxiliary charger, owing to the convexity of ergotropy~\cite{Francica2017,Bernards2019}.  

%\begin{figure}
%    \centering
%    \includegraphics[width=\linewidth]{idea.pdf}
%    \caption{Reconciling charging protocols involving (a) measuring the auxiliary charger, and (b) tracing out (discarding) the auxiliary charger to harness the daemonic advantage.}
%    \label{fig:idea}
%\end{figure}

Our paper demonstrates that harnessing the daemonic advantage due to measurement on the auxiliary charger while defining the charging protocol via tracing out the same is possible, and repeated application of the charging cycle can access the daemonic sector of work extraction.  For this, we determine the time-maximized ergotropy $\mathcal{E}^{\max}$ of $\rho_b$, occurring at a time $t=\tau$, and specifically ask whether the minimum daemonic ergotropy, $\overline{\mathcal{E}}_{\min}(\tau)$, minimized over all possible projection measurement on the auxiliary charger,  at $t=\tau$ can be reached by $\mathcal{E}^{\max}$. We define the \emph{daemonic gap} as $\overline{\mathcal{E}}_{\min}(\tau)-\mathcal{E}^{\max}$, and show that a gapless charging cycle can be designed when a harmonic mode is considered as the battery, interacting with an auxiliary qubit charger via the Jaynes-Cummings interaction~\cite{Jaynes1963,Cummings1965,Larson2021}. We explore the gaplessness of the designed charging cycle for different passive initial states of the harmonic mode battery, including the ground state, truncated mixture of low-lying states~\cite{Lvovsky2002,Kaneda2016,Davis2022}, and canonical thermal states. We further generalize the bipartite battery-charger model to a quantum battery constituted of multiple harmonic modes, interacting with different pairs of energy levels in an auxiliary qudit charger with the ground state in common, through the Jaynes-Cummings interaction, and demonstrate construction of gapless charging cycle for a double-mode quantum battery and a qutrit charger, for different passive battery-initial states.

Further, we define the \emph{daemonic band} at $t=\tau$ as the difference between the minimum and maximum possible daemonic ergotropy, i.e., $\overline{\mathcal{E}}_{\max}(\tau)-\overline{\mathcal{E}}_{\min}(\tau)$, where both the maximization and minimization are performed over the complete set of projection measurements on the auxiliary charger. We show, in the case of the single-mode battery and the qubit charger, that choosing the ground state of the harmonic mode as the initial state results in a collapse of the band. On the other hand, for other initial states including the truncated mixture of low-lying states and the canonical thermal states, the daemonic band is negligible compared to the maximum amount of energy that can be stored in the battery, and   repeated application of the designed charging cycle can achieve full charging of the battery, thereby accessing the full daemonic band. We highlight the advantage of the proposed repeated application of the charging cycle specifically in the practical problem of simultaneously charging components of a composite battery (cf.~\cite{Binder2015,barra2022,Dias2024}). Particularly, we show, using a double-mode battery and a qutrit charger, that both the modes can not be simultaneously charged using the designed cycle, starting from their respective ground states. However, applying the charging cycle multiple times following our prescription leads to simultaneous charging of the two modes.  

The rest of the paper is organized as follows. In Sec.~\ref{sec:competition}, we formally introduce the charging cycle (Sec.~\ref{subsec:cc}), and discuss the concept of the daemonic gap and the daemonic band (Sec.~\ref{subsec:daemonic_gap}). The battery-charger model constituted of a single-mode battery and a qubit charger interacting via the Jaynes-Cummings interaction is introduced in Sec.~\ref{sec:single_mode}, and realization of gapless charging starting from the ground state of the battery is demonstrated in Sec.~\ref{subsec:ground_state}.  The role of the initial battery state in the gapless charging is discussed in Sec.~\ref{subsec:battery_initial_state}, while accessing the daemonic band and full charging of the battery using sequential application of the designed cycle is demonstrated  in Sec.~\ref{subsec:repeated_charging}. The advantage of the proposed scheme in simultaneously charging multi-mode quantum batteries with a auxiliary qudit charger is demonstrated in Sec.~\ref{sec:double_mode}. Sec.~\ref{sec:conclusion} contains concluding remarks and outlook.

\begin{figure}
    \centering
    \includegraphics[width=\linewidth]{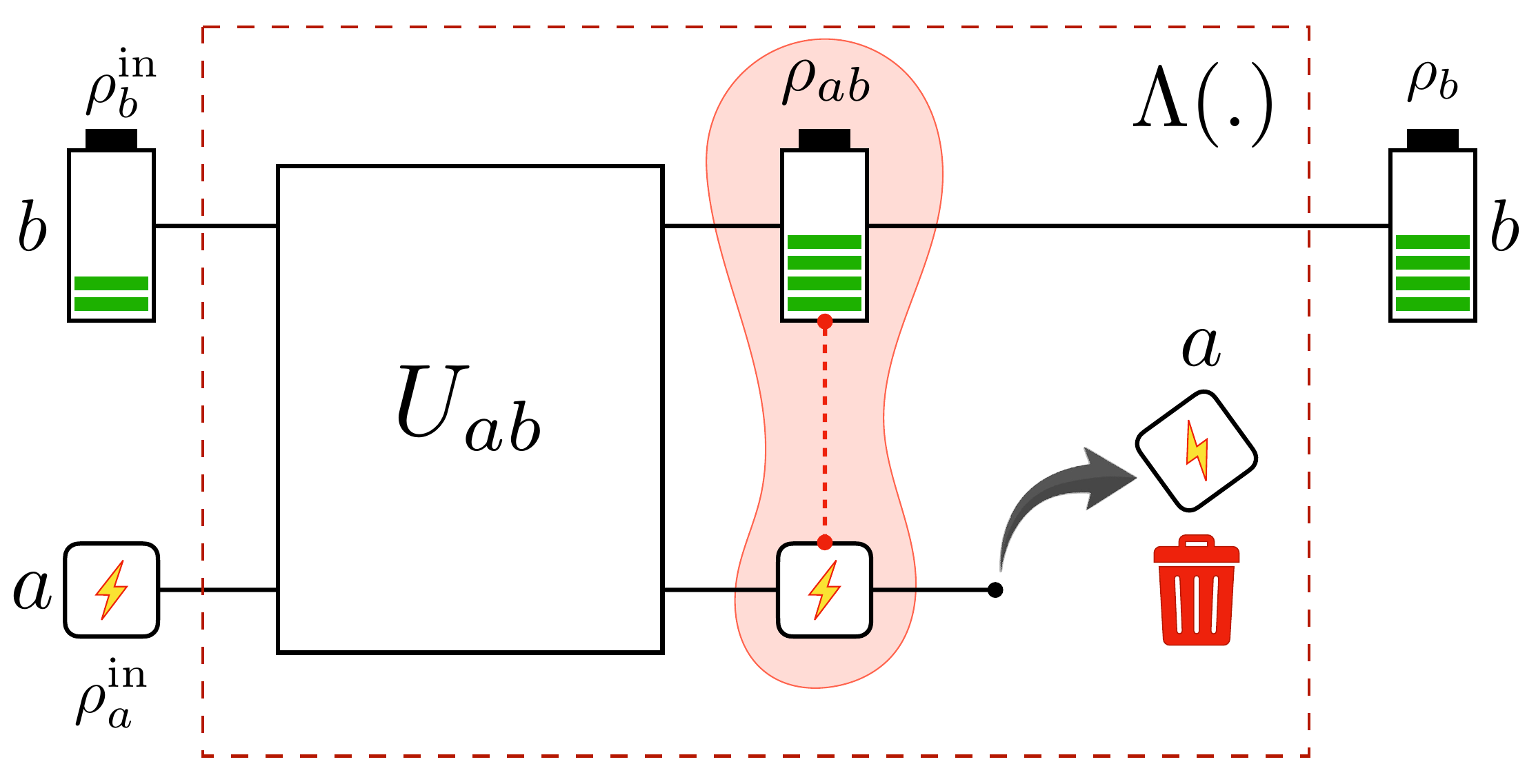}
    \caption{\textbf{Charging cycle.} The battery (charger) is initialized in the state $\rho_b^{\text{in}}$ ($\rho_a^{\text{in}}$). The operation $U_{ab}$ entangles the battery and the charger, resulting in the state $\rho_{ab}$. This is followed by a measurement and subsequent discarding of the charger, leaving the battery in the state $\rho_b$ with ergotropy $\mathcal{E}$.}
    \label{fig:charging_cycle}
\end{figure}

\section{Charging via a single auxiliary}
\label{sec:competition}

In this section, we formally introduce the charging cycle for a bipartite battery-charger setup. We also define the daemonic gap, and discuss its relevance in connection to the performance of the charging cycle.  

\subsection{Charging cycle}
\label{subsec:cc}

Consider a $d_b$-dimensional quantum battery $b$ described by the Hamiltonian $H_b$ satisfying the eigenvalue equation $H_b\ket{n}=\omega_n\ket{n}$, connected to a $d_a$-dimensional \emph{auxiliary} charger $a$ represented by the Hamiltonian $H_a$ satisfying $H_a\ket{e_n}=\nu_n\ket{e_n}$. The battery and the charger interacts via an interaction Hamiltonian $H_{ab}$, such that the total Hamiltonian describing the battery-charger duo is given by $H=H_a+H_b+H_{ab}$.  At $t=0$, the interaction $H_{ab}$ is absent, such that the battery-charger Hamiltonian is given by $H_a+H_b$, and the battery-charger duo is prepared in the state $\rho_{ab}^{\text{in}}=\rho_a^{\text{in}}\otimes\rho_b^{\text{in}}$. Here, $\rho_a^{\text{in}}$ is a judiciously chosen state of the charger, and  $\rho_{b}^{\text{in}}$ is typically a completely \emph{passive} state of the battery~\cite{Alicki2013,Campaioli2018,Campaioli2024,Pusz1978,Singh2021,Allahverdyan2004}, having the form 
\begin{eqnarray}
    \rho_b^{\text{in}}=\sum_{n=0}^{d_b-1} r_{n}^{(0)}\ket{n}\bra{n},
    \label{eq:battery_initial_state}
\end{eqnarray}
with $r_n^{(0)}$ being real, $0\leq r_n^{(0)}\leq 1$ $\forall n$, and  $r_0^{(0)}\geq r_1^{(0)}\geq \cdots \geq r_{d_b-1}^{(0)}\geq r_{d_{b}}^{(0)}$.

At $t>0$, the battery undergoes an operation $\Lambda$ given by (see Fig.~\ref{fig:charging_cycle})
\begin{eqnarray}
    \rho_b^{\text{in}}\rightarrow\rho_b&=&\Lambda(\rho_b^{\text{in}})=\text{Tr}_a\left[U_{ab}\rho_{ab}^{\text{in}}U_{ab}^\dagger\right],
    \label{eq:cc}
\end{eqnarray}
where $\rho_b$ is typically a mixed state, and $U_{ab}$ is an entangling unitary operation generated by $H$, i.e., $U_{ab}=\exp\left[-\text{i}H t\right]$,  resulting in an entangled state $\rho_{ab}=U_{ab}\rho_{ab}^{\text{in}}U_{ab}^\dagger$.  starting from $\rho_{ab}^{\text{in}}$. From $\rho_b$, a passive state $\varrho_b=\sum_{k=0}^{d_b-1}q_k\ket{k}\bra{k}$ can be constructed, where $\{q_k\}$ are the eigenvalues of $\rho_b$ arranged in descending order (i.e., $q_{k}\geq q_{k+1}$, $0\leq k\leq d_b-1$), and $\{\ket{k}\}$ are the energy eigenvectors of $H_b$ having energy eigenvalues $E_k$ such that $E_k\leq E_{k+1}$, $0\leq k\leq d_b-1$. The ergotropy~\cite{Allahverdyan2004} $\mathcal{E}(t)$ of $\rho_b$ at an arbitrary time $t$ quantifies the maximum attainable work from $\rho_b$ with respect to $H_b$ under cyclic unitaries, and is given by \begin{eqnarray}
    \mathcal{E}(t)&=&\text{Tr}\left[H_{b}\left\{\rho_{b}-\varrho_{b}\right\}\right]. 
\end{eqnarray}
The operation $\Lambda$ (Eq.~(\ref{eq:cc})), referred to as the \emph{charging cycle}, charges the battery $b$ if non-zero ergotropy can be obtained from the state $\rho_b$. Assuming that $\mathcal{E}(t)$ is maximum at $t=\tau$, 
\begin{eqnarray}
    \mathcal{E}^{\max}=\mathcal{E}(\tau)=\max_{t}\mathcal{E}(t). 
\end{eqnarray}

\subsection{Daemonic gap and daemonic band}
\label{subsec:daemonic_gap} 
  
Each of the battery-state $\rho_b$ at an arbitrary time $t$ corresponds to an infinite set of ensembles $\left\{\rho_b^k=\text{Tr}_a\left[\tilde{\rho}_{ab}^k\right],p^k\right\}$. Each such ensemble is generated by a member $M_a$ of the infinite set of possible projection measurements $\mathcal{M}_a=\left\{M_a\right\}$ on $a$, and is constituted of the post-measured battery states $\tilde{\rho}_{b}^k$ obtained from the post-measured battery-charger state $\tilde{\rho}_{ab}^k$ corresponding to the measurement outcome $k$, occurring with the probability $p^k$, such that $\rho_b=\sum_kp^k\rho_b^k$.  Defining $M_a$ by the projection operators $\{P_a^{k}=\ket{k}\bra{k}\}$ corresponding to the measurement outcomes $k$,  $\tilde{\rho}_{ab}^k=\left(P_a^k\rho_{ab}P_a^k\right)/p^k$, and $p^k=\text{Tr}\left[P_a^k\rho_{ab}P_a^k\right]$. Note that both $\rho_b^{k}$ and $p^{k}$ are functions of $t$ in general, which we discard from the notations for brevity. It is also worthwhile to note that one may consider all possible measurements including the POVMs while defining the daemonic ergotropy. However, in this paper, we focus only on the projection measurements.

At an arbitrary time $t$, the daemonic ergotropy~\cite{Francica2017,Bernards2019,Morrone2023,Barra2024,Kua2025} corresponding to each measurement $M_a$  is defined as
\begin{eqnarray}
    \overline{\mathcal{E}}(t,M_a)=\sum_{k}p^{k}\mathcal{E}^{k}(t,M_a),
\end{eqnarray}
with $\mathcal{E}^{k}(t,M_a)$ being the ergotropy~\cite{Allahverdyan2004} of  $\rho_b^{k}$ at $t$. The convexity of ergotropy implies $\overline{\mathcal{E}}(t,M_a) \geq \mathcal{E}(t)$  $\forall t$ and $\forall M_a$, while the advantage due to access to the  outcomes corresponding to the measurement $M_a$ at $t$ is given by 
\begin{eqnarray}
    \mathcal{A}(t,M_a)=\overline{\mathcal{E}}(t,M_a)-\mathcal{E}(t).
\end{eqnarray}

We define the \emph{daemonic gap} as the \emph{daemonic advantage over $\mathcal{E}(t)$ minimized over $\mathcal{M}_a$ at $t$}, given by 
\begin{eqnarray}
    \mathcal{L}(t)&=&\min_{\mathcal{M}_a}\mathcal{A}(t,M_a)=\overline{\mathcal{E}}_{\min}(t)-\mathcal{E}(t), 
\end{eqnarray}
where we have defined $\overline{\mathcal{E}}_{\min}(t)=\min_{\mathcal{M}_a}\overline{\mathcal{E}}(t,M_a)$.  Given a fixed passive initial battery state $\rho_b^{\text{in}}$ (Eq.~(\ref{eq:battery_initial_state})), a fixed initial charger state $\rho_a^{\text{in}}$, and a fixed battery-charger Hamiltonian $H=H_a+H_b+H_{ab}$ (i.e., a specific $U_{ab}$), we are specifically interested at the time $t=\tau$ at which $\mathcal{E}(t)$ is maximum (see Sec.~\ref{subsec:cc}), such that $\mathcal{A}(\tau,M_a)$ quantifies the \emph{daemonic gap corresponding to the measurement $M_a$ at $t=\tau$}. Subsequently, $\mathcal{L}(\tau)$ quantifies the \emph{overall minimum daemonic gap} at $t=\tau$. Note that $\mathcal{L}(\tau)>0$ implies that for the charging cycle defined in Sec.~\ref{subsec:cc}, all projection measurements $M_a$ on $a$ provides a non-zero daemonic advantage at $t=\tau$, implying an irreducible daemonic gap. On the other hand, $\mathcal{L}(\tau)=0$ corresponds to a \emph{gapless} cycle, for which $\mathcal{E}^{\max}$ equals to the daemonic ergotropy corresponding to a subset $\mathcal{M}_a^0=\{M_a^0\}\subset\mathcal{M}_a$ of measurements with cardinality at least $1$.

It is straightforward to see that at $t=\tau$, $0\leq \mathcal{L}(\tau)\leq\mathcal{U}(\tau)$, where 
\begin{eqnarray}
    \mathcal{U}(t)=\max_{\mathcal{M}_a}\mathcal{A}(t,M_a)= \overline{\mathcal{E}}_{\max}(t)-\mathcal{E}(t),
    \label{eq:daemonic_gain}
\end{eqnarray}
is the maximum \emph{daemonic gain}~\cite{Francica2017} at $t$, with $\overline{\mathcal{E}}_{\max}(t)=\max_{\mathcal{M}_a}\overline{\mathcal{E}}(t,M_a)$. This motivates the definition of the \emph{daemonic band}, quantifying the range of work that can be accessed only by allowing all possible projection measurements on the charger, given by 
\begin{eqnarray}
    \Delta(\tau)&=&\mathcal{U}(\tau)-\mathcal{L}(\tau)
    =\overline{\mathcal{E}}_{\max}(\tau)- \overline{\mathcal{E}}_{\min}(\tau). 
\end{eqnarray}
Note that a collapse of the daemonic band at $t=\tau$, i.e., $\Delta(\tau)=0$ suggests $\overline{\mathcal{E}}_{\max}(\tau)=\overline{\mathcal{E}}_{\min}(\tau)$, and subsequently a  measurement-independent $\overline{\mathcal{E}}(\tau)$.

\section{Closing the gap in a single-mode battery}
\label{sec:single_mode}

We now consider a battery modeled by a single harmonic mode of energy $\omega$ with $d_b$ levels, such that  
\begin{eqnarray}
    H_b=\omega\sum_{n=0}^{d_b-1}n\ket{n}\bra{n}, 
    \label{eq:battery_hamiltonian}
\end{eqnarray} 
where $\omega>0$, and $\{\ket{n}:n=0,1,2,\cdots,d_b-1\}$ can be identified as the Fock states.  
The charger is  represented by a qubit ($d_a=2$), with the Hamiltonian 
\begin{eqnarray}
    H_a=\nu\ket{e_1}\bra{e_1},
    \label{eq:qubit_charger_hamiltonian}
\end{eqnarray}
where we have assumed the energy of the ground state $\ket{e_0}$ of the qubit to be zero ($\nu_0=0$, $\nu_1=\nu$, see Sec.~\ref{subsec:cc}). The battery and the charger interact via the Jaynes-Cummings interaction~\cite{Jaynes1963,Cummings1965,Larson2021}, given by  
\begin{eqnarray}
\label{eq:single_mode_battery_charger_interaction}
    H_{ab}&=&g\left(\ket{e_0}\bra{e_1}\otimes \mathcal{O}^\dagger_{b}+\ket{e_1}\bra{e_0}\otimes \mathcal{O}_{b}\right),
\end{eqnarray}
where $g$ denotes the strength of the interaction, and  
\begin{eqnarray}
    \mathcal{O}_{b}^\dagger&=&\sum_{n=0}^{d_b-2} \sqrt{n+1}\ket{n+1}\bra{n},\nonumber\\
    \mathcal{O}_{b}&=&\sum_{n=1}^{d_b-1} \sqrt{n}\ket{n-1}\bra{n},
    \label{eq:raising_lowering_operator}
\end{eqnarray}
are the raising and lowering operators for the harmonic mode. The unitary operator $U_{ab}$ in the charging cycle $\Lambda$ (Eq.~(\ref{eq:cc})) is now generated by $H=H_a+H_b+H_{ab}$, where $H_b$, $H_a$, and $H_{ab}$ are given by Eqs.~(\ref{eq:battery_hamiltonian}), (\ref{eq:qubit_charger_hamiltonian}), and (\ref{eq:single_mode_battery_charger_interaction}) respectively. The following proposition describes the output battery state $\rho_b$ obtained from $\Lambda$.  

\PROP{I} Starting from  $\rho_a^{\text{in}}=\ket{e_1}\bra{e_1}$ and $\rho_b^{\text{in}}$ as in Eq.~(\ref{eq:battery_initial_state}),  $\Lambda$ (Eq.~(\ref{eq:cc})) leads to a diagonal $\rho_b$ in the Fock state basis. 

\begin{proof} 
The time evolution $U_{ab}$ of $\rho_{a}^{\text{in}}\otimes\rho_b^{\text{in}}$ leads to
\begin{eqnarray}
    \rho_{ab}&=&\sum_{n=1}^{d_b-1}r_{n-1}^{(0)}\Big[\Big\{A_n(t)  \ket{n,e_0}\bra{n,e_0}\nonumber\\
    &&+B_n(t)\ket{n-1,e_1}\bra{n-1,e_1}\Big\}\nonumber\\
    &&+\left(C_n(t)\ket{n,e_0}\bra{n-1,e_1}+\text{h.c.}\right)\Big]\nonumber\\
    &&+r_{d_b-1}^{(0)}\ket{d_b-1,e_1}\bra{d_b-1,e_1},
    \label{eq:single_mode_time_evolved_state}
\end{eqnarray}
with
\begin{eqnarray}
    \Omega_n&=&\sqrt{g^2n+\delta^2/4},\nonumber\\
    A_n(t)&=& \left[g^2n\sin^2\Omega_n t\right]/\Omega_n^2,\nonumber\\
    B_n(t) &=& \left[4g^2n\cos^2\Omega_nt+\delta^2\right]/4\Omega_n^2,\nonumber\\
    C_n(t) &=& \frac{g\sqrt{n}}{2\Omega_n^2}\left[\delta \sin^2(\Omega_nt)-\text{i} \Omega_n \sin(2\Omega_nt)\right],
\end{eqnarray}
where $\delta=\omega-\nu$ is the detuning. Subsequently,
\begin{eqnarray}
    \rho_b^{(1)} &=& r_{d_b-1}^{(0)}\ket{d_b-1}\bra{d_b-1}+\sum_{n=1}^{d_b-1} r_{n-1}^{(0)}\Big[A_n(t) \ket{n}\bra{n}\nonumber\\&&+B_n(t) \ket{n-1}\bra{n-1}\Big]\nonumber\\&=& \sum_{n=0}^{d_b-1} r_n^{(1)} \ket{n}\bra{n},
    \label{eq:battery_state_diagonal}
\end{eqnarray}
is diagonal in the energy eigenbasis of $H_b$, having eigenvalues
\begin{eqnarray}
    r_0^{(1)}(t)&=&r_0^{(0)}B_1(t),\nonumber\\ 
    r_i^{(1)}(t)&=&r_{i-1}^{(0)}A_i(t)+r_i^{(0)}B_{i+1}(t),\nonumber\\
    r_{d_b-1}^{(1)}(t)&=&r_{d_b-1}^{(0)}+r_{d_b-2}^{(0)}A_{d_b-1}(t), 
    \label{eq:eigenvalues_rho_b_delta}
\end{eqnarray}
for $1\leq i\leq d_b-2$, with explicit time-dependence, where the subscript as well as the superscript ``$(1)$" indicates that the charging cycle has been applied only once. Hence the proof.
\end{proof}

\subsection{Gapless charging from ground state}
\label{subsec:ground_state}

Proposition I leads to the following results with the ground state as the initial battery-state.

\PROP{II} With $\rho_b^{\text{in}}=\ket{0}\bra{0}$, $\mathcal{E}(t)$ is maximized at $\tau=(2\ell+1)\pi/2\Omega_1$, $\ell=0,1,2,\cdots$.

\begin{proof}
    With $\rho_b^{\text{in}}=\ket{0}\bra{0}$ (i.e., $r_0^{(0)}=1$, $r_{n>0}^{(0)}=0$),
    \begin{eqnarray}
        \rho_b&=&A_1(t)\ket{0}\bra{0}+B_1(t)\ket{1}\bra{1},  
    \end{eqnarray}
    leading to
    \begin{eqnarray}
    \label{eq:ergo_vacuum}
    \mathcal{E}(t)&=&\left\{\begin{array}{l}
       \omega \{A_1(t)-B_1(t)\} \text{ for } A_1(t)>B_1(t),   \\
         0, \text{  otherwise}, 
    \end{array}\right. \end{eqnarray}
    which  is non-zero in the interval \begin{eqnarray}
    t&\in& \left(\frac{\alpha+2\pi \ell}{\Omega_1},\frac{\pi-\alpha+2\pi \ell}{\Omega_1}\right)\nonumber\\&&\cup \left(\frac{\pi+\alpha+2\pi \ell}{\Omega_1},\frac{2\pi-\alpha+2\pi \ell}{\Omega_1}\right),\nonumber\\&\equiv& \Delta t,
    \label{single_cycle_interval}
    \end{eqnarray} 
    assuming its maximum value $\omega$ at $\tau=(2\ell+1)\pi/2\Omega_1$ $\forall \ell=0,1,2,\cdots$. Here,  $\alpha=\cos^{-1}\sqrt{(4g^2-\delta^2)/8g^2}$. Hence the proof.
\end{proof}

\PROP{III} With $\rho_b^{\text{in}}=\ket{0}\bra{0}$, the daemonic band collapses  $\forall t$.  

\begin{proof}
To prove this, consider projection measurement on $a$ in the arbitrary complete orthonormal basis $\{\ket{\xi_+},\ket{\xi_-}\}$, corresponding to the measurement outcomes $k=\pm 1$ occurring with probabilities $p^{\pm 1}$,  where 
\begin{eqnarray}
    \ket{\xi_+}&=&\cos\frac{\alpha}{2}\ket{e_0}+\text{e}^{i\gamma}\sin\frac{\alpha}{2}\ket{e_1},\nonumber\\
    \ket{\xi_-}&=&\sin\frac{\alpha}{2}\ket{e_0}-\text{e}^{i\gamma}\cos\frac{\alpha}{2}\ket{e_1},
    \label{eq:measurement_basis}
\end{eqnarray}
with $(\alpha,\gamma)\in\mathbb{R}$, $0\leq\alpha\leq \pi$, and $0\leq \gamma\leq 2\pi$. With $\rho_{b}^{\text{in}}=\ket{0}\bra{0}$, projection measurement on the charger defined by the basis in Eq.~(\ref{eq:measurement_basis}), when the battery-charger duo is in the state $\rho_{ab}$ (Eq.~(\ref{eq:single_mode_time_evolved_state})),  leads to the post-measured ensemble
\begin{eqnarray}
     \rho_{b}^{+1}&=&\frac{1}{p_{(+1)}}\Bigg[A_1(t) \cos^2\frac{\alpha}{2} \ket{1}\bra{1}+B_1(t) \sin^2 \frac{\alpha}{2}\ket{0}\bra{0}\nonumber\\
    &&+\left(\frac{1}{2}\text{e}^{i\gamma}\sin \alpha C_n(t)\ket{1}\bra{0}+\text{h.c.}\right)\Bigg]\nonumber \\
     \rho_{b}^{-1}&=&\frac{1}{p_{(-1)}}\Bigg[A_1(t) \sin^2\frac{\alpha}{2} \ket{1}\bra{1}+B_1(t) \cos^2 \frac{\alpha}{2}\ket{0}\bra{0}\nonumber\\
    &&-\left(\frac{1}{2}\text{e}^{i\gamma}\sin \alpha C_n(t)\ket{1}\bra{0}+\text{h.c.}\right)\Bigg]
\end{eqnarray}
on the battery, with probabilities
\begin{eqnarray}
    p^{+1}&=&A_1(t)\cos^2 \alpha/2+B_1(t)\sin^2 \alpha/2,\nonumber\\
    p^{-1}&=&A_1(t)\sin^2 \alpha/2+B_1(t)\cos^2 \alpha/2,
\end{eqnarray}
respectively. Subsequently, 
\begin{eqnarray}
\label{eq:daemonic_vacuum}
    \overline{\mathcal{E}}(t)=\omega A_1(t), 
\end{eqnarray}
which is independent of $\alpha,\gamma$ $\forall t$, implying $\Delta(t)=0$. Hence the proof.
\end{proof}

\begin{figure}
    \centering
    \includegraphics[width=0.9\linewidth]{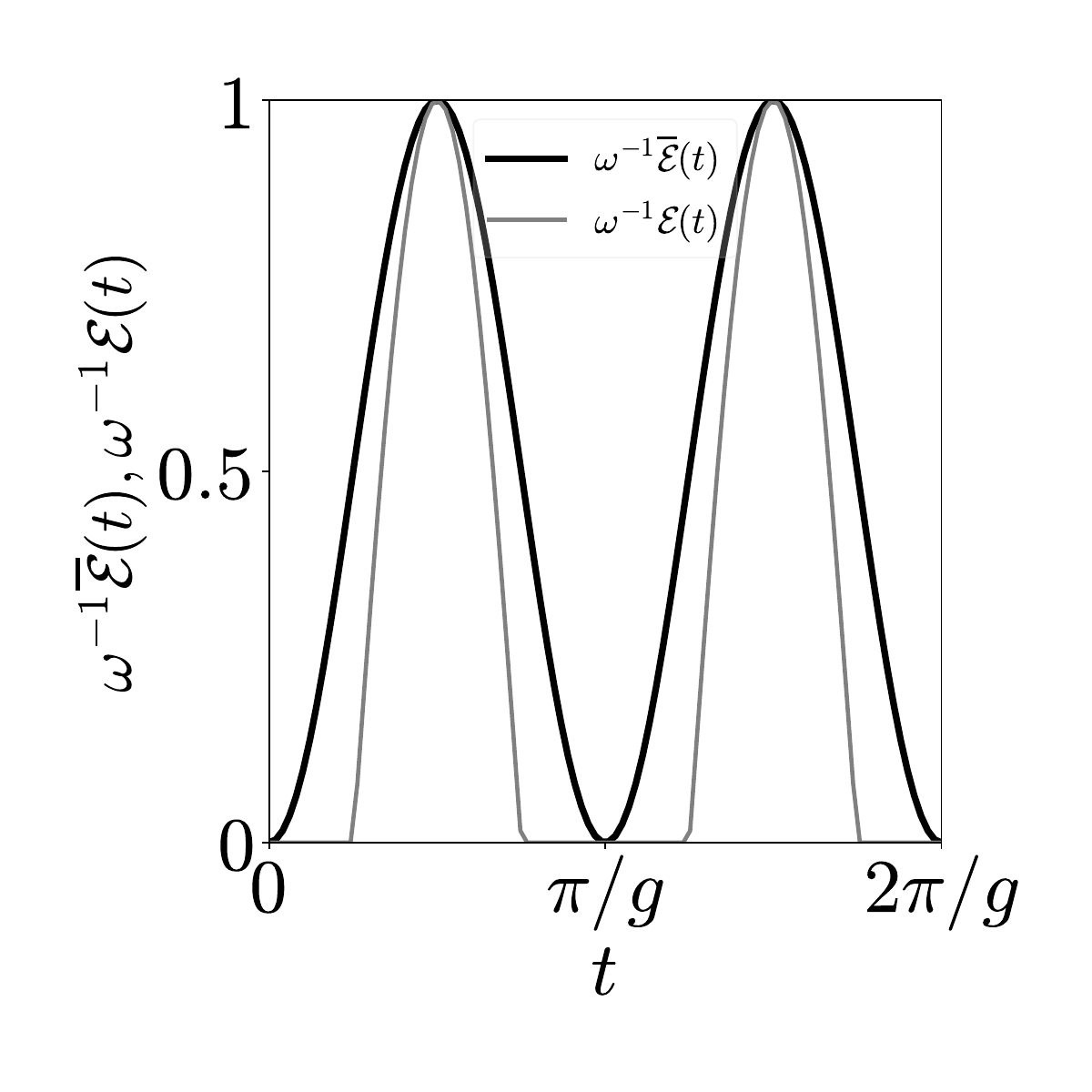}
    \caption{Variations of $\omega^{-1}\overline{\mathcal{E}}(t)$ and $\omega^{-1}\mathcal{E}(t)$ against $t$.  }
    \label{fig:single_mode_ground_state}
\end{figure}

\PROP{IV} For $\rho_b^{\text{in}}=\ket{0}\bra{0}$, $\mathcal{L}(\tau)=0$ when $\delta=0$, while with $\delta\neq 0$, $\mathcal{L}(\tau)=\omega \delta^2/4\Omega_1^2$. 
 
\begin{proof}
    Using Eq.~(\ref{eq:daemonic_vacuum}), $\overline{\mathcal{E}}(\tau)=\omega=\mathcal{E}^{\max}$, implying $\mathcal{L}(\tau)=0$ for $\delta=0$. However, for $\delta\neq0$, $\overline{\mathcal{E}}(\tau)=g^2 \omega/\Omega_1^2$, and $\mathcal{E}^{\max}=\omega(4g^2-\delta^2) /4\Omega_1^2$, where $\tau=(2\ell+1)\pi/2\Omega_1$. In this case,  $\mathcal{L}(\tau)=\omega\delta^2/4\Omega_1^2>0$ for $\delta>0$. Hence the proof. 
\end{proof}

\begin{figure*}
    \centering
    \includegraphics[width=0.8\linewidth]{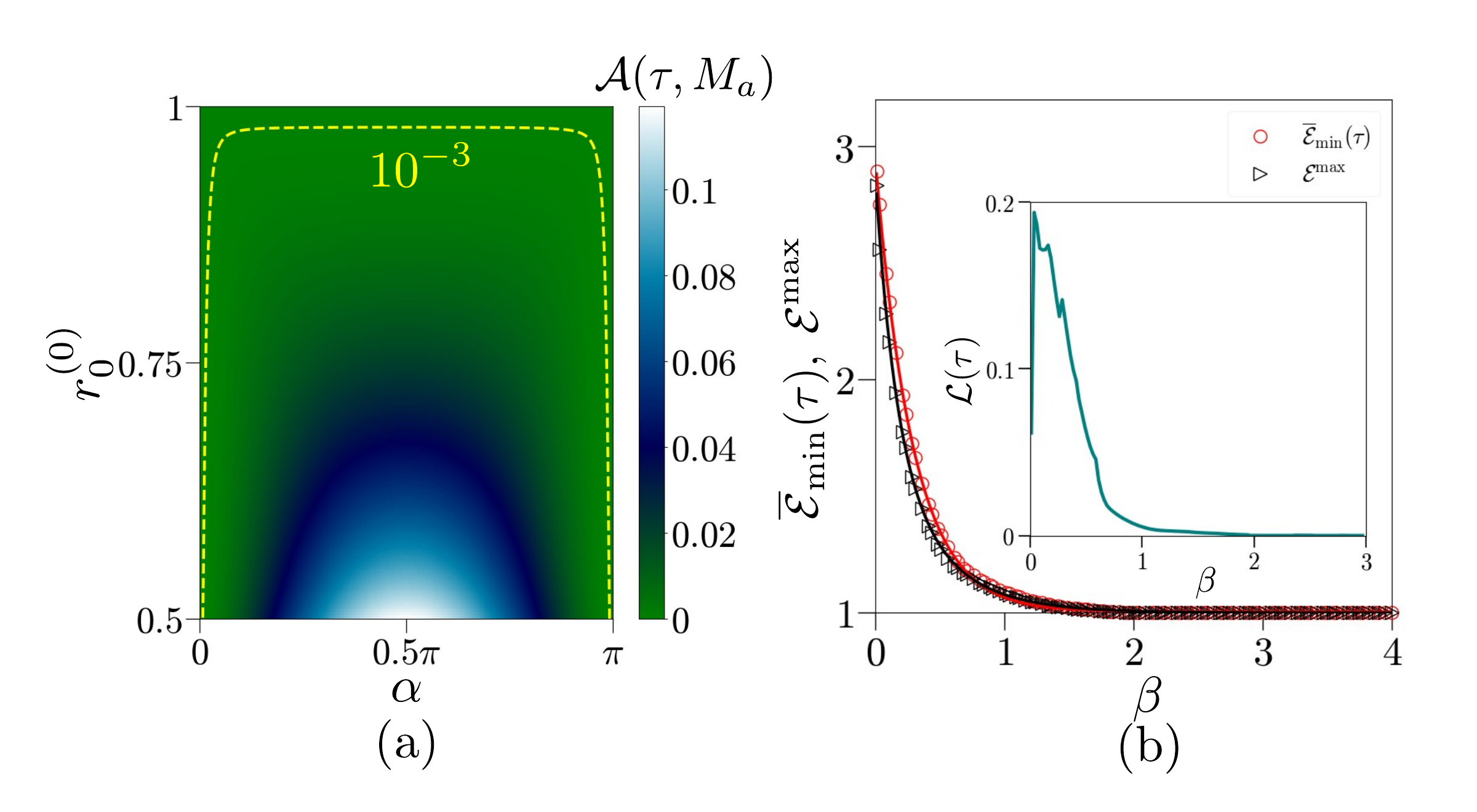}
    \caption{(a) Landscape of $\mathcal{A}(\tau,M_a)$ on the $(r_0^{(0)},\alpha)$ plane, indicating that the minimum occurs at $\alpha=0,\pi$ $\forall r_0^{(0)}$. (b) Variations of  $\overline{\mathcal{E}}_{\min}(\tau)$, $\mathcal{E}^{\max}$, and $\mathcal{L}(\tau)$ (inset) as functions of $\beta$. The trends of $\mathcal{E}^{\max}$ and $\overline{\mathcal{E}}_{\min}(\tau)$  against $\beta$ are fitted to Eqs.~(\ref{eq:fitted_1}) and (\ref{eq:fitted_2}) respectively, with the fitting parameters given by $c_0=1.77$,  $c_1=3.75$, $\overline{c}_0=1.94$, $\overline{c}_1=3.442$. To generate both plots (a) and (b), we have fixed $\omega=1$, $d_b=11$ and $g=1$. }
    \label{fig:single_mode_truncated_states}
\end{figure*}

\noindent In Fig.~\ref{fig:single_mode_ground_state}, we plot $\overline{\mathcal{E}}(t)$ and $\mathcal{E}(t)$ as functions of $t$ in the interval $[0,2\pi/g]$ setting $\delta=0$ and $g=1$, demonstrating Propositions II-IV.

\subsection{Dependence on different initial battery states}
\label{subsec:battery_initial_state}

We now investigate the daemonic gap corresponding to  $\Lambda$ for two specific initial states, namely, the truncated mixture of low-lying states, and the canonical thermal states, as discussed below.     

\subsubsection{Truncated mixture of low-lying states} 

These states have the form
\begin{eqnarray}
    \rho_{b,d}^{\text{tr}} &=&\sum_{n=0}^{d-1}r_{n}^{(0)}\ket{n}\bra{n},
    \label{eq:truncated_states}
\end{eqnarray}
where $r_0^{(0)}\geq r_1^{(0)}\geq \cdots \geq r_{d-2}^{(0)}\geq r_{d-1}^{(0)}$, and $\sum_{n=0}^{d-1}r_n^{(0)}=1$, with $d\ll d_b$, potentially resulting from imperfect heralding~\cite{Lvovsky2002,Kaneda2016,Davis2022}. Note that $\rho_{b,1}^{\text{tr}}=\ket{0}\bra{0}$ (see Sec.~\ref{subsec:ground_state}), while in this Section, we specifically concentrate on the cases of $d=2$ and $3$ motivated by the literature~\cite{Lvovsky2002,Kaneda2016,Davis2022}.

The two-level truncated mixture $\rho_{b,2}^{\text{tr}}$ is passive if $r_0^{(0)}\geq 1/2$, $r_1^{(0)}=1-r_0^{(0)}\leq 1/2$. Using Proposition I, the battery state at an arbitrary time $t$ can be written as
\begin{eqnarray}
\rho_{b}^{(1)}&=&\sum_{n=0}^2r_n^{(1)}\ket{n}\bra{n}, 
\end{eqnarray}
with 
\begin{eqnarray}
    r_0^{(1)}&=&r_0^{(0)}B_1(t),\nonumber\\
    r_1^{(1)}&=&r_0^{(0)}A_1(t)+\big(1-r_0^{(0)}\big)B_2(t),\nonumber\\
    r_2^{(1)}&=&(1-r_0^{(0)})A_2(t).
\end{eqnarray}   

While determining $\mathcal{E}(t)$ analytically is non-trivial due to the dependence on the value of $r_0^{(0)}$, we approach the investigation of $\mathcal{L}(\tau)$ numerically.  Our  analysis indicates that for $\delta=0$,  $\mathcal{E}(t)$ is maximum at $\tau=(2\ell+1)\pi/2g$, $\ell=0,1,2,\cdots$ for all $r_0^{(0)}\geq 1/2$, where $\mathcal{E}^{\max}=\omega$, and  $r_0^{(1)}=0$, $r_1^{(1)}+r_2^{(1)}=1$ with both $r_1^{(1)},r_2^{(1)}>0$. On the other hand, for $\delta\neq 0$, we choose the maximum possible value of $\delta$ to be $10\%$ of the characteristic frequency $\omega_b$ of the battery, i.e., $0< \delta\leq 10^{-1}\omega$. Initial battery states of the form $\rho_{b,2}^{\text{tr}}$ 
requires the time evolution to continue till $\tau=(2\ell+1)\pi/2\Omega_1$, as suggested by our numerical analysis, to achieve $\mathcal{E}^{\max}$ in the chosen range of $\delta$.

It is also cumbersome to analytically determine $\mathcal{L}(\tau)$ as a function of ($\alpha$, $\gamma$) for arbitrary values of $r_{0}^{(0)}$. However, our numerical analysis reveals that $\mathcal{A}(\tau,M_a)$ is independent of $\gamma$ irrespective of whether $\delta=0$ or $\neq 0$. For $\delta=0$, the daemonic ergotropy at $t=\tau$ is minimum at $\alpha=0,\pi$, i.e., corresponding to measurement in the energy eigenbasis $\{\ket{e_0},\ket{e_1}\}$ of $a$, providing $\mathcal{L}(\tau)=0$ (according to our numerical accuracy, we consider $\mathcal{L}(\tau)\leq 10^{-3}$ to be $\mathcal{L}(\tau)=0$). See Fig.~\ref{fig:single_mode_truncated_states}(a) for an illustration. However, for $\delta\neq 0$, there exists initial battery states  $\rho_{b,2}^{\text{tr}}$ leading to $\mathcal{L}(\tau)>0$. This can be demonstrated with the example given by $r_0^{(0)}=0.9$, $r_1^{(0)}=0.1$, for which $\mathcal{L}(\tau)=2\times 10^{-3}$, where the minimization takes place at $\alpha=0$. 

%\PROP{VI} 

%\begin{proof}
%     Hence the proof. 
%\end{proof}

We next consider $\rho_{b,3}^{\text{tr}}$, for which passivity is ensured by
\begin{eqnarray}
    r_0^{(0)}&\geq& 1/3,\text{ and }
    r_0^{(0)} \geq r_1^{(0)}\geq (1-r_0^{(0)})/2,  
    \label{eq:three_level_mixture_condition}
\end{eqnarray}
leading to (see Proposition I)
\begin{eqnarray}
    \rho_b^{(1)}&=&\sum_{n=0}^3r_n^{(1)}\ket{n}\bra{n},
\end{eqnarray}
with
\begin{eqnarray}
    r_0^{(1)}(t)&=&r_0^{(0)}B_1(t),\nonumber\\ 
    r_1^{(1)}(t)&=&r_{0}^{(0)}A_1(t)+r_1^{(0)}B_{2}(t),\nonumber\\
    r_2^{(1)}(t)&=&r_{1}^{(0)}A_2(t)+r_2^{(0)}B_{3}(t),\nonumber\\   r_3^{(1)}(t)&=&r_{2}^{(0)}A_3(t).  
\end{eqnarray} 
In contrast to the case of $\rho_{b,2}^{\text{tr}}$, in this case, there exists $\rho_{b,3}^{\text{tr}}$ which, when used as $\rho_b^{\text{in}}$, results in $\mathcal{L}(\tau)>0$ for the entire range $0\leq\delta\leq 10^{-1}\omega$. For demonstration,  at $\delta=0$, consider the example of $\rho_{b,3}^{\text{tr}}$ with $r_0^{(0)}=0.43$, and $r_1^{(0)}=0.42$, for which  $\tau=1.46\pi$, and $\mathcal{L}(\tau)=1.1\times 10^{-2}$, with the minimization occurring at $\alpha=0.12\pi$. On the other hand, for $\delta=10^{-1}\omega$, $\tau=1.46\pi$, $\mathcal{L}(\tau)=1.4\times 10^{-2}$, and $\alpha=0.14\pi$ corresponding to the minimum $\mathcal{L}(\tau)$.

\subsubsection{Canonical thermal state} 

We now consider the canonical thermal states, which are obtained when the resonator mode equilibrates in contact with a thermal reservoir of absolute temperature $T$, and are  given by 
\begin{eqnarray}        \rho_b^{\text{th}}= Z^{-1}\exp\left[-\beta H_{b}\right],
\end{eqnarray}
where $Z=\text{Tr}\left[\exp{\left(-\beta H_{b}\right)}\right]$ is the partition function, and $\beta=T^{-1}$ (assuming $k_B=1$), such that 
\begin{eqnarray}
    r_{n}^{(0)}=Z^{-1}\exp{[-\beta \omega n]},
    \label{eq:boltzmann}
\end{eqnarray}
and the condition of passivity of $\rho_b^{\text{in}} (=\rho_b^{\text{th}})$ is satisfied. 
Note that $\rho_{b,d}^{\text{tr}}$ with an appropriate $d$ can also be obtained with the populations following Eq.~(\ref{eq:boltzmann}) when $\beta$ is non-zero and finite. 

%\RESULT{III} With $\rho_b^{\text{in}}=\rho_b^{\text{th}}$ and for $\delta=0$, $\Lambda$ is gapless for $1.65\lesssim\beta\leq\infty$.

Similar to the case of $\rho_{b,d}^{\text{tr}}$, analytical calculation for $\rho_b^{\text{in}}=\rho_b^{\text{th}}$ is non-trivial. However, our detailed numerical analysis suggests that  for $\delta=0$, $\tau$ depends on $\beta$, while $\mathcal{E}^{\max}=\omega$. Further, $\mathcal{A}(\tau,M_a)$ is independent of $\gamma$ $\forall$ $\beta$, and is also independent of $\alpha$ for $1.65\lesssim\beta\leq\infty$, while $\mathcal{L}(\tau)=0$ in this range of $\beta$ (see Fig.~\ref{fig:single_mode_truncated_states} (b)). On the other hand, outside the range $1.65\lesssim\beta\leq\infty$, $\overline{\mathcal{E}}_{\min}(\tau)$ deviates from $\mathcal{E}^{\max}$, as illustrated in Fig.~\ref{fig:single_mode_truncated_states}(b), leading to a non-zero $\mathcal{L}(\tau)$. The overall variations of $\mathcal{E}^{\max}$ and $\overline{\mathcal{E}}_{\min}(\tau)$ with $\beta$ is, however, similar, given by 
\begin{eqnarray}
         \label{eq:fitted_1}
        \mathcal{E}^{\max}&=&1+c_0\text{e}^{-c_1\beta},\\
        \overline{\mathcal{E}}_{\min}(\tau)&=&1+\overline{c}_0\text{e}^{-\overline{c}_1\beta},
        \label{eq:fitted_2}
\end{eqnarray}
where both $\overline{\mathcal{E}}_{\min}(\tau),\mathcal{E}^{\max}\rightarrow 1$ as $\beta\rightarrow\infty$.

\noindent\textbf{Note.} It is worthwhile to note that  while specific values of $\omega,\nu$, and $g$ are chosen to obtain these results,  the findings  remain qualitatively unchanged even with varying system parameters.

\begin{figure*}
    \centering
    \includegraphics[width=\linewidth]{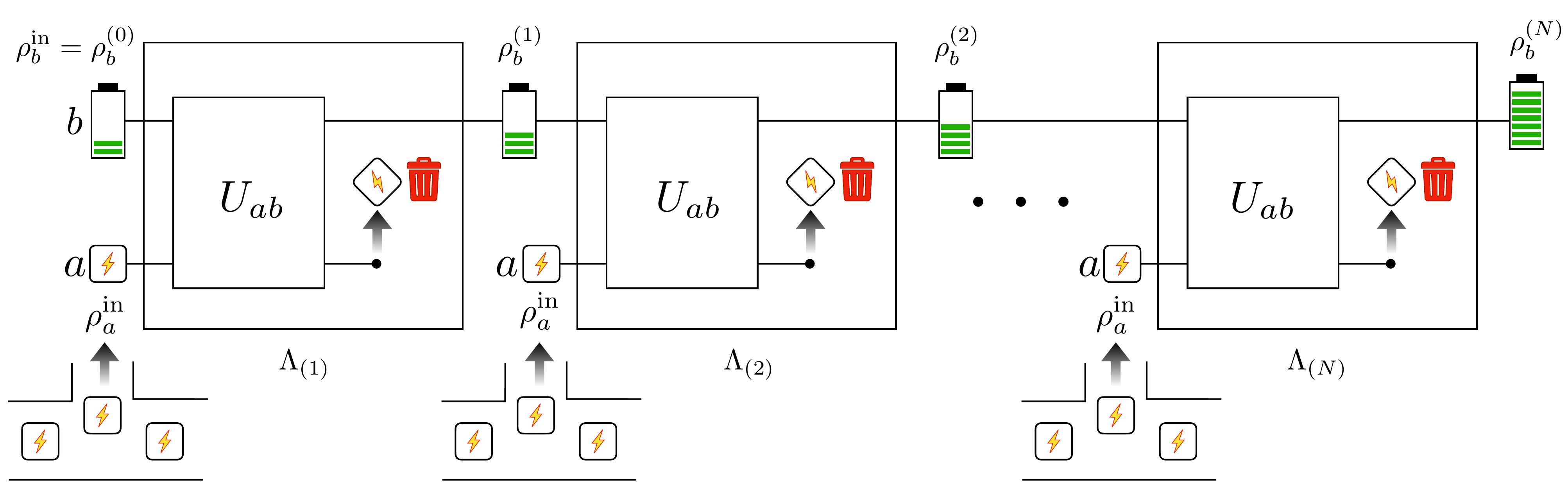}
    \caption{Schematic representation of repeated application of $\Lambda(.)$ on the battery-charger setup. See Sec.~\ref{subsec:repeated_charging}.}
    \label{fig:repeated}
\end{figure*}

\subsection{Repeating charging cycles}
\label{subsec:repeated_charging}

For a specific initial state $\rho_{b}^{\text{in}}$, $\Lambda$ can be repeated for $N$ rounds of charging, given by 
\begin{eqnarray}
    \boldsymbol{\Lambda}=\Lambda_{(N)}\circ\cdots\circ\Lambda_{(m)}\circ\cdots\circ\Lambda_{(2)}\circ\Lambda_{(1)},
\end{eqnarray}
with
\begin{eqnarray}
    \rho_b^{(N)}=\boldsymbol{\Lambda}\big(\rho_b^{(0)}\big),
\end{eqnarray}
with each round $\Lambda_{(m)}$ $(1\leq m\leq N-1)$ defined as in Eq.~(\ref{eq:cc}), providing the maximum ergotropy $\mathcal{E}^{\max}_{(m)}$ at a time $\tau_{(m)}$ as defined in Sec.~\ref{subsec:cc}, till a full charging, i.e., $\mathcal{E}^{\max}_{(N)}= (d_b-1)\omega$, takes place. Here, $\rho_{b}^{\text{in}}=\rho_b^{(0)}$ serves as the initial state of $\Lambda_{(1)}$, providing the output $\rho_b^{(1)}$, discussed in Secs.~\ref{subsec:ground_state} and \ref{subsec:battery_initial_state}. The output $\rho_b^{(m)}$ of the $m$th charging round  serves as the input for $\Lambda_{(m+1)}$ (see Fig.~\ref{fig:repeated}). We denote the number of rounds of $\Lambda$ required to fill the daemonic gap to be $N_{\text{gap}}$. 
%while the number of rounds in which the battery cross the daemonic band is represented by $N_{\text{band}}$. 
Note that for $\delta=0$ and $\rho_b^{\text{in}}=\ket{0}\bra{0}$ and $\rho_{b,2}^{\text{tr}}$, $N_{\text{gap}}=1$ (see Secs.~\ref{subsec:ground_state} and \ref{subsec:battery_initial_state}), while for other initial states, $N_{\text{gap}}$ may be $>1$. We investigate the $N_{\text{gap}}$ for different initial states in this section.

\noindent\textbf{Note.} Allowing repetition of $\Lambda$, we assume availability of a set of qubits, each initialized in the state $\ket{e_1}$, such that the charger can be attached to the battery at the beginning of every cycle. See Fig.~\ref{fig:repeated}. 

The following proposition extends Proposition I to the case of repeated application of the charging cycle. 

\PROP{V} Starting from  $\rho_a^{\text{in}}=\ket{e_1}\bra{e_1}$ and $\rho_b^{\text{in}}$ as in Eq.~(\ref{eq:battery_initial_state}),  application of $\Lambda$ (Eq.~(\ref{eq:cc})) $m$ times leads to a diagonal $\rho_b^{(m)}$ in the Fock state basis.

\begin{proof}
    The proof is straightforward, leading to \begin{eqnarray}
    \rho_b^{(m)}=\sum_{n=0}^N r_n^{(m)} \ket{n}\bra{n},
    \end{eqnarray}
    with 
    \begin{eqnarray}
    r_0^{(m)}(t)&=&r_0^{(m-1)}B_1(t),\nonumber\\ 
    r_i^{(m)}(t)&=&r_{i-1}^{(m-1)}A_i(t)+r_i^{(m-1)}B_{i+1}(t),\nonumber\\
    r_{d_b-1}^{(m)}(t)&=&r_{d_b-1}^{(m-1)}+r_{d_b-2}^{(m-1)}A_{d_b-1}(t),  
   \label{eq:eigenvalues_rho_b}
   \end{eqnarray}
   for $1\leq i\leq d_b-2$, and $m\geq 1$. 
\end{proof}

\begin{figure*}
    \centering
    \includegraphics[width=0.9\linewidth]{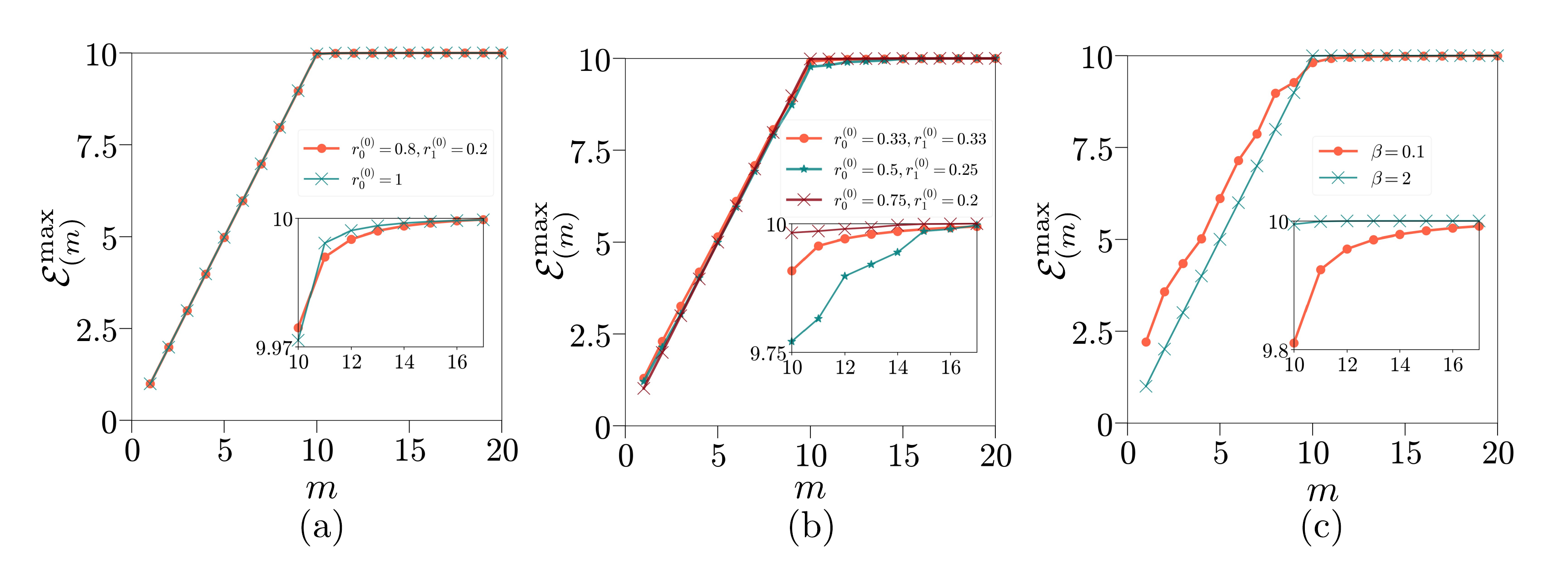}
    \caption{\textbf{Full charging by repeating charging cycle.}  Variations of $\mathcal{E}^{\max}$ with increasing charging rounds $m$, where the duration of the time evolution in the $m$th round is $\tau_{(m)}$. The chosen initial states are (a) ground state of $b$ and truncated mixtures of two lowest lying states, (b) truncated mixtures of three levels, and (c) canonical thermal states. For (a) and (b), we fix $\delta=0.1$, while for (c), $\delta=0$. To generate the plots, we have fixed $\omega=1$, $d_b=11$ and $g=1$.}
    \label{fig:single_mode_repeated}
\end{figure*}

We first consider $\rho_b^{(0)}=\rho_b^{\text{in}}=\ket{0}\bra{0}$, which leads to the following results. 

\PROP{VI} For $\rho_{b}^{\text{in}}=\rho_b^{(0)}=\ket{0}\bra{0}$ and $\delta=0$, $N_{\text{gap}}=1$, while full charging is achieved in $N=d_b-1$ cycles.  

%\PROP{VII} For $\rho_{b}^{\text{in}}=\rho_b^{(0)}=\ket{0}\bra{0}$ and $\delta=0$, the single-mode battery is . 

\begin{proof}
The proof of $N_{\text{gap}}=1$  follows directly follows from Propositions III and IV. Further, complete charging of the single-mode battery $b$ in tune ($\delta=0$) with the qubit charger $a$ corresponds to the occupation of the state $\ket{d_b-1}$. With $\rho_b^{(0)}=\ket{0}\bra{0}$ and $\delta=0$, the charging cycle can be applied $m$ times such that (see proof of Proposition II-IV for the first round) 
\begin{eqnarray}
    \rho^{(m)}_b&=&B_m(t)\ket{m-1}\bra{m-1}+A_m(t)\ket{m}\bra{m}.
\end{eqnarray}
Maximizing $\mathcal{E}_{(m)}(t)$ w.r.t. $t$  leads to  $\mathcal{E}^{\max}_{(m)}=m\omega$ and $\rho_b^{(m)}=\ket{m}\bra{m}$ at $\tau_{(m)}=(2\ell+1)\pi/2\Omega_{m}$ (i.e., $r_m^{(m)}=1$ and $r_{n\neq m}^{(m)}=0$), provided $\rho_{b}^{(m-1)}=\ket{m-1}\bra{m-1}$ occurring at $\tau_{(m-1)}$ is the input to $\Lambda_{(m)}$. The state $\ket{d_b-1}$, therefore, is occupied after $N=d_b-1$ rounds.
\end{proof}

While obtaining closed-form expressions is cumbersome for $0<\delta\leq 10^{-1}\omega$, our numerical investigation reveals that similar to the case of $\delta=0$, $\tau_{(m)}=(2\ell+1)\pi/2\Omega_m$, $\ell=0,1,2,\cdots$. However, in contrast to the case of $\delta=0$, $\rho_b^{(m)}$ corresponds to $r_{n\leq m}^{(m)}>0$, and $r_{n> m}^{(m)}=0$ when $0<\delta\leq 10^{-1}\omega$, and full charging requires more than $d_b-1$ steps, as demonstrated in Fig.~\ref{fig:single_mode_repeated}(a).

Next, we apply this for $\rho_b^{\text{in}}=\rho_{b,d}^{\text{tr}}$ with $d\ll d_b$, focusing specifically on  $d=2$. Our numerical analysis for $\delta=0$ leads to $\tau_{(m)}=(2\ell+1)\pi/2\sqrt{m}g$, and 
\begin{eqnarray}
    \rho^{(m)}_b&=&r_{m-1}^{(m-1)}B_m(t)\ket{m-1}\bra{m-1}\nonumber\\&&+\bigg[r_{m-1}^{(m-1)}A_m(t)+r_m^{(m-1)}B_{m+1}(t)\bigg]\ket{m}\bra{m}\nonumber\\
    &&+r_m^{(m-1)}A_{m+1}(t)\ket{m+1}\bra{m+1},
\end{eqnarray}
for $\forall m< d_b-1$, while after the application of $\Lambda_{(d_b-1)}$, 
\small 
\begin{eqnarray}
\rho^{(d_b-1)}_b&=&r_{d_b-2}^{(d_b-2)}B_{d_b-1}(t)\ket{d_b-2}\bra{d_b-2}\nonumber \\    &&+\bigg[r_{d_b-2}^{(d_b-2)}A_{d_b-1}(t)+r_{d_b-1}^{(d_b-2)}\bigg]\ket{d_b-1}\bra{d_b-1},\nonumber\\ 
\end{eqnarray}\normalsize
which, at $\tau_{(d_b-1)}$, is $\rho_{b}^{(d_b-1)}=\ket{d_b-1}\bra{d_b-1}$. Therefore, similar to the case of the ground state as the initial battery state, for $\rho_{b}^{\text{in}}=\rho_b^{(0)}=\rho_{b,2}^{\text{tr}}$ with $\delta=0$, maximum charging occurs in $N=d_b-1$ steps.

In contrast, in the case of $0<\delta\leq 10^{-1}\omega$, starting from $\rho_{b,2}^{\text{tr}}$, one may require $>d_b-1$ charging rounds to achieve full charging, as also demonstrated in Fig.~\ref{fig:single_mode_repeated}(a). Similar observation is made in the cases of $\rho_{b,3}^{\text{tr}}$ and $\rho_b^{\text{th}}$ as initial states, as evident from Figs.~\ref{fig:single_mode_repeated}(b) and (c) respectively. 

Our numerical calculation further demonstrates that the daemonic gap $\Delta(\tau)$ (where $\tau\equiv\tau_{(1)}$ by definition, see Sec.~\ref{subsec:daemonic_gap}), for all the initial states considered in this paper, is very small compared to the maximum energy that can be stored in the battery. Therefore, the fact that repeating the charging cycle achieves full charging of the battery for all of these initial states automatically imply a full access of the daemonic band.

\section{Parallel charging multi-mode batteries}
\label{sec:double_mode}

\begin{figure}
    \centering
    \includegraphics[width=0.9\linewidth]{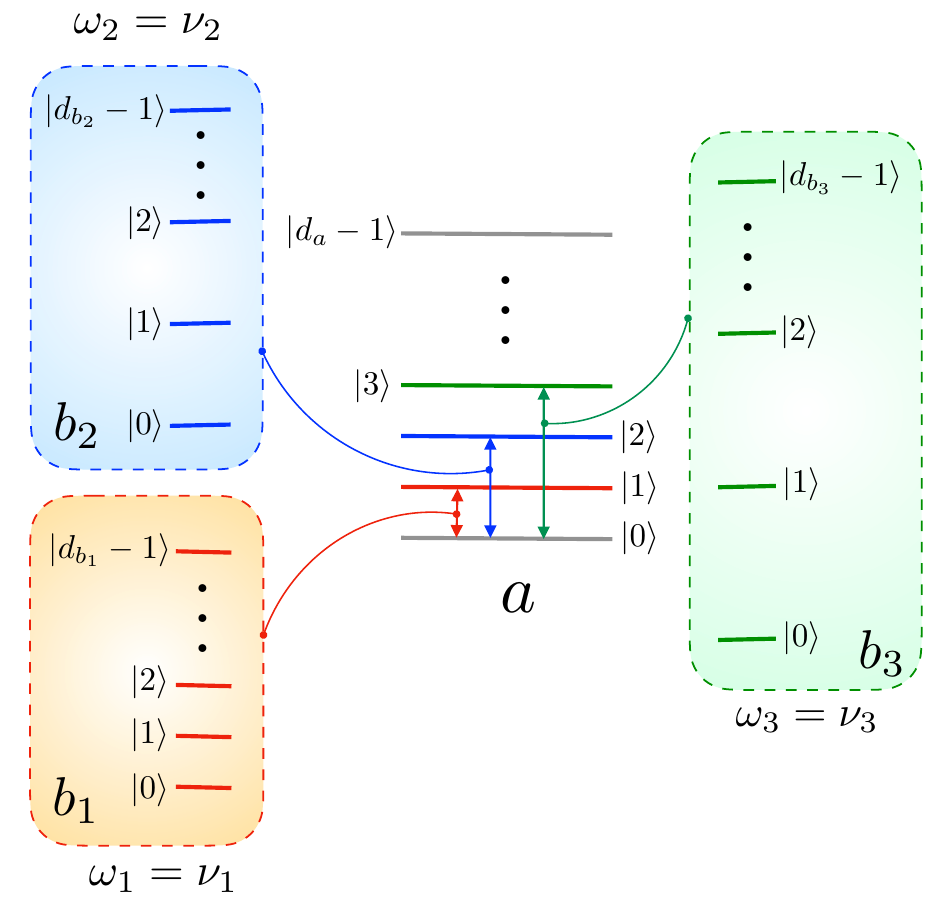}
    \caption{Illustration of the battery-charger interaction Hamiltonian $H_{ab}$, where three modes $b_1$, $b_2$, and $b_3$ are illustrated. Each mode $b_i$ connects to the transition between the levels $\{\ket{e_0},\ket{e_i}\}$ of the charger $a$, corresponding to an energy $\nu_i$, via an interaction Hamiltonian of strength $g$ having the form as given in Eq.~(\ref{eq:single_mode_battery_charger_interaction}) (with $\ket{e_1}$ replaced by $\ket{e_i}$), such that the total interaction Hamiltonian takes the form of Eq.~(\ref{eq:double_mode_battery_charger_interaction}). Here, we have assumed detuning $\delta_i=\omega_{i}-\nu_i=0$ $\forall i=1,2,\cdots,d_{a}-1$.} 
    \label{fig:hab}
\end{figure}

We now generalize the design of the charging cycle for a collective battery $b$ of $d_a-1$ modes  $\{b_i;i=1,2,\cdots,d_a-1\}$. Each mode $b_i$ with $d_{b_i}$ energy levels is described by the Hamiltonian 
\begin{eqnarray}
    H_{b_i}=\omega_i\sum_{n_i=0}^{d_{b_i}-1}n_i\ket{n_i}\bra{n_i},
\end{eqnarray}
while the total battery Hamiltonian is given by $H_b=\sum_{i=1}^{d_a-1}H_{b_i}$. We consider the battery modes to be \emph{non-degenerate}, i.e., $\omega_{i}\neq\omega_{j}$, $i,j=1,2,\cdots,d_a-1$, $j\neq i$. On the other hand,  the auxiliary charger $a$ is a fully accessible qudit of dimension $d_a$, given by the Hamiltonian
\begin{eqnarray}
    H_a&=&\sum_{i=0}^{d_a-1}\nu_i\ket{e_i}\bra{e_i},
    \label{eq:qudit_charger_hamiltonian}
\end{eqnarray}
where $H_a\ket{e_i}=\nu_i\ket{e_i}$ for $i=0,1,\cdots,d_a$, with $\nu_0=0$, $\nu_{i>0}>0$, and $\nu_{i+1}>\nu_i$ $\forall i$. The battery and the charger interact via the Hamiltonian 
\begin{eqnarray}
\label{eq:double_mode_battery_charger_interaction}
    H_{ab}&=&g\sum_{i=1}^{d_a-1}\left(\ket{e_0}\bra{e_i}\otimes \mathcal{O}^\dagger_{b_i}+\ket{e_i}\bra{e_0}\mathcal{O}_{b_i}\right),
\end{eqnarray}
where each $b_i$ is interacting with the pair of levels $\{\ket{e_0},\ket{e_i}\}$ of the charger qudit, with a strength  $g$ of interaction that is identical for all such pairs (see Fig.~\ref{fig:hab}). The unitary operation $U_{ab}$ involved in the charging cycle $\Lambda$ (see Sec.~\ref{subsec:cc}) is generated by the total Hamiltonian $H=H_a+H_b+H_{ab}$. The detuning corresponding to each mode can be defined as $\delta_i=\omega_i-\nu_i$, which we set to zero $\forall i$ to keep the calculations uncluttered.

We assume that at $t=0$, the battery-charger duo is prepared in the state $\rho_{ab}^{\text{in}}=\rho_a^{\text{{in}}}\otimes \rho_{b}^{\text{in}}$, where $\rho_a^{\text{in}}=\ket{\psi}\bra{\psi}$ 
is a state in the excited state subspace of the auxiliary qudit charger, and $\rho_b^{\text{in}}=\bigotimes_{i=1}^{d_a}\rho_{b_i}^{\text{in}}$. Here,  
\begin{eqnarray}
    \rho_{b_i}^{\text{in}}=\sum_{n_i=0}^{d_{b_i}-1} r_{n_i}^{(0)}\ket{n_i}\bra{n_i},
    \label{eq:individual_battery_initial_state}
\end{eqnarray}
is a passive state of the mode $b_i$ with additional constraints on $\{r_{n_i}^{(0)}\}$ such that $\rho_b^{\text{in}}$ is also a passive state of the battery collective $b$. The charging cycle $\Lambda$ at $t>0$ leads to $\rho_{b}=\Lambda(\rho_b^{\text{in}})$, providing ergotropy $\mathcal{E}(t)$, while the state $\rho_{b_i}=\text{Tr}_{b\backslash b_i}\left[\rho_b\right]$ corresponding to the mode $b_i$ provides ergotropy $\mathcal{E}_{b_i}(t)$. A \emph{simultaneous (parallel) charging} of all modes $b_i$ along with the battery collective $b$ is said to have happened if
\begin{eqnarray}
    \label{eq:simultaneous_charging}   \mathcal{E}_b>0\text{ and }\mathcal{E}_{b_i}>0,\;\forall i=1,2,\cdots,d_a-1.
\end{eqnarray}
Note further that the definitions of daemonic gap and daemonic band can be extended to the collective battery as well as the individual battery modes. In this paper, however, we focus on only the daemonic gap and the daemonic band corresponding to the collective battery $b$. 

\subsection{Charging a double-mode battery}
\label{subsec:double_mode_battery}

For demonstration, we consider the special case of $d_a=3$, and assume that the initial state of $a$ is a generic state
\begin{eqnarray}
    \ket{\psi}=\cos\frac{\theta}{2}\ket{e_1}+\text{e}^{\text{i}\phi}\sin\frac{\theta}{2} \ket{e_2}, 
\end{eqnarray}
in the subspace spanned by $\ket{e_1}$ and $\ket{e_2}$, where $0\leq \theta\leq \pi$, $0\leq \phi\leq 2\pi$. With this, the battery $b$ can be charged in a gapless fashion starting from its ground state, as shown in the following proposition.

\PROP{VII} For a double-mode battery $b=\{b_1,b_2\}$ with $\rho_{b_i}^{\text{in}}=\ket{0}\bra{0}$ for $i=1,2$,   $\mathcal{L}(\tau)=0$.

\begin{proof}
We follow the convention $\ket{n_1n_2\cdots n_i\cdots}$ while writing multi-mode states, discarding the subscript ``$i$" labeling the modes for brevity. Considering the ground state as the initial state for both qudits, i.e.,  $\rho_b^{\text{in}}=\ket{00}\bra{00}$,  
\begin{eqnarray}
    \rho_{b}^{(1)}&=&\sin^2 gt\Bigg\{\cos^2\frac{\theta}{2} \ket{10}\bra{10}+\sin^2\frac{\theta}{2}\ket{01}\bra{01} \nonumber\\&&+\frac{1}{2}\sin \theta \bigg(\text{e}^{i\phi}\ket{01}\bra{10}+\text{h.c.}\bigg)\Bigg\}+\cos^2 gt \ket{00}\bra{00},\nonumber\\
    \label{eq:collective_state_first_round}
\end{eqnarray} 
leading to 
\begin{eqnarray}
    \mathcal{E}^{(1)}_b(t)&=&\sin^2gt\left(\omega_{1} \cos^2\frac{\theta}{2}+\omega_{2}\sin^2\frac{\theta}{2}\right)\nonumber\\&&-\omega_{2}\cos^2gt,
\end{eqnarray}
for $t\in\Delta t$, and 
\begin{eqnarray}
    \mathcal{E}^{(1)}_b(t)&=&\sin^2gt\left(\omega_{1} \cos^2\frac{\theta}{2}+\omega_{2}\sin^2\frac{\theta}{2}\right)\nonumber\\&&-\omega_{2}\sin^2gt,
\end{eqnarray}
for $t\notin\Delta t$ and $\omega_2<\omega_1$, where
\begin{eqnarray}
    \Delta t=\left(\frac{\pi}{4g}+\frac{\ell\pi}{g},\frac{3\pi}{4g}+\frac{\ell\pi}{g}\right), 
\end{eqnarray}
with $\ell$ being positive integers. Clearly,  $\mathcal{E}^{(1)}_b(t)$ attains the maximum  value $\omega_1\cos^2(\theta/2)+\omega_2 \sin^2(\theta/2)$  at $\tau=(2\ell+1)\pi/2g$, $\ell=0,1,2,\cdots$.  

On the other hand, considering measurement in energy eigenbasis $\{\ket{e_0},\ket{e_1},\ket{e_1}\}$, 
$\rho_{b}^{(1),k}$ corresponding to the measurement outcomes $k=0,1,2$ can be calculated as 
\begin{eqnarray}
    \rho_{b}^{(1),0}&=&\cos^2\frac{\theta}{2} \ket{10}\bra{10}+\sin^2\frac{\theta}{2}\ket{01}\bra{01}\nonumber\\&&+\frac{1}{2}\sin \theta (\text{e}^{i\phi}\ket{01}\bra{10}+\text{h.c.}), 
    \label{eq:collective_post_measured_states_0}
\end{eqnarray}
and 
\begin{eqnarray}
    \rho_b^{(1),1}&=& \rho_b^{(1),2}=\ket{00}\bra{00},
    \label{eq:collective_post_measured_states_12}
\end{eqnarray}
occurring with the probabilities $p_b^{0}= \sin^2 gt$, $p_b^{1}=\cos^2 \frac{\theta}{2} \cos^2 gt$, and $p_b^{2}=\sin^2 \frac{\theta}{2} \cos^2 gt$
respectively. This results in
\begin{eqnarray}
    \overline{\mathcal{E}}_b(t)&=&\sin^2gt\left(\omega_{1} \cos^2\frac{\theta}{2}+\omega_{2}\sin^2\frac{\theta}{2}\right),
\end{eqnarray}
which, at $t=\tau$, becomes $\omega_1\cos^2(\theta/2)+\omega_2 \sin^2(\theta/2)$, thereby equal to $\mathcal{E}^{\max}$. Hence the proof.
\end{proof}

It is now logical to ask whether both modes $b_1$ and $b_2$ are simultaneously charged at $t=\tau$.  The following Proposition proves that it is not possible at any $t$ by a single application of $\Lambda$, but can be possible by applying the charging cycle at least twice. 

\PROP{VIII} The charging cycle $\Lambda$ has to be applied at least twice for parallel charging both $b_1$ and $b_2$, starting from $\rho_{b_i}^{\text{in}}=\ket{0}\bra{0}$, $i=1,2$, provided the initial state of the charger is given by $\ket{\psi}$ with the state parameter $\theta$ satisfying $\frac{1}{2}\sin^2 \theta \geq \sin^4 \theta/2$ and $\frac{1}{2}\sin^2 \theta \geq \cos^4 \theta/2$ simultaneously. 

\begin{proof}
From Eq.~(\ref{eq:collective_state_first_round}), we obtain 
\begin{eqnarray}
    \rho_{b_1}^{(1)}&=&\sin^2 gt \cos^2\frac{\theta}{2} \ket{1}\bra{1}\nonumber\\&&+(\sin^2 gt\sin^2\frac{\theta}{2}+\cos^2 gt )\ket{0}\bra{0},\nonumber\\
    \rho_{b_2}^{(1)}&=&\sin^2 gt \sin^2\frac{\theta}{2} \ket{1}\bra{1}\nonumber\\&&+(\sin^2 gt\cos^2\frac{\theta}{2}+\cos^2 gt )\ket{0}\bra{0}.    
\end{eqnarray}
Note that at all $t$, the condition for non-passivity of $\rho_{b_1}^{(1)}$ and $\rho_{b_2}^{(1)}$ are given respectively by 
$\cot^2 gt < \cos\theta$ and $\cot^2 gt > \cos\theta$, which can not be simultaneously satisfied, implying that  $\mathcal{E}_{b_i}\ngtr 0$ $\forall i=1,2$ at any $t$, including $\tau_{(1)}$.  

However, a second application of $\Lambda$ leads to
\begin{widetext}
\begin{eqnarray}
    \rho_b^{(2)} &=&\sin^22\sqrt{2}gt\bigg[\cos^4\frac{\theta}{2}\ket{20}\bra{20}+\sin^4\frac{\theta}{2}\ket{02}\bra{02}
    +\frac{1}{2}\sin^2 \theta\ket{11}\bra{11}\bigg]+\cos^2 2\sqrt{2}gt \bigg[\cos^2\frac{\theta}{2} \ket{10}\bra{10}\nonumber\\
    &&+ \sin^2\frac{\theta}{2}\ket{01}\bra{01}\bigg]+\bigg[\sin^22\sqrt{2}gt\bigg\{\frac{\text{e}^{2i\phi}}{4}\sin^2 \theta\ket{02}\bra{20}+\sqrt{2}\text{e}^{i\phi}\sin\theta\bigg(\cos^2\frac{\theta}{2}\ket{11}\bra{20}\nonumber\\&&+\text{e}^{-2i\phi}\sin^2\frac{\theta}{2}\ket{11}\bra{02}\bigg)\bigg\}+\frac{\text{e}^{i\phi}}{2}\sin \theta\cos^2 2\sqrt{2}gt \ket{01}\bra{10}\bigg],
    \label{eq:second_cycle_double_mode}
\end{eqnarray}
\end{widetext}
%where we have chosen $\theta=\pi/2$ and $\phi=0$ for brevity. 
and subsequently,  
%\begin{eqnarray}
%    \rho_{b_i}^{(2)}&=&\frac{\sin^2 2\sqrt{2}gt}{4}\ket{2}\bra{2}+\frac{1}{2}\ket{1}\bra{1}\nonumber\\&&+\frac{1+\cos^2 2\sqrt{2}gt}{4}\ket{0}\bra{0},;i=1,2, 
%\end{eqnarray}
\small 
\begin{eqnarray}
    \rho_{b_1}^{(2)}&=&\sin^2 2\sqrt{2}gt \cos^4 \frac{\theta}{2}\ket{2}\bra{2}+\Big(\frac{1}{2}\sin^2 2\sqrt{2}gt \sin^2 \theta\nonumber \\
    &&+\cos^2 2\sqrt{2}gt \cos^2 \frac{\theta}{2}\Big)\ket{1}\bra{1}+\Big(\sin^2 2\sqrt{2}gt \sin^4 \frac{\theta}{2}\nonumber\\&&+\cos^2 2\sqrt{2}gt \sin^2 \frac{\theta}{2}\Big)\ket{0}\bra{0},\nonumber \\
     \rho_{b_2}^{(2)}&=&\sin^2 2\sqrt{2}gt \sin^4 \frac{\theta}{2}\ket{2}\bra{2}+\Big(\frac{1}{2}\sin^2 2\sqrt{2}gt \sin^2 \theta\nonumber \\
    &&+\cos^2 2\sqrt{2}gt \sin^2 \frac{\theta}{2}\Big)\ket{1}\bra{1}+\Big(\sin^2 2\sqrt{2}gt \cos^4 \frac{\theta}{2}\nonumber\\&&+\cos^2 2\sqrt{2}gt \cos^2 \frac{\theta}{2}\Big)\ket{0}\bra{0}
    \label{eq:simultaneous_charging_cycle_2}
\end{eqnarray}\normalsize
From Eq.(\ref{eq:second_cycle_double_mode}), it follows that the collective charging reaches its maximum at $\tau_{(2)}=(2\ell+1)\pi/4\sqrt{2}g$. Substituting in Eq.~(\ref{eq:simultaneous_charging_cycle_2}) implies that simultaneous charging of both modes requires $\frac{1}{2}\sin^2 \theta \geq \sin^4 \theta/2$ and $\frac{1}{2}\sin^2 \theta \geq \cos^4 \theta/2$ simultaneously, defining the range of $\theta$, and thereby the set of initial charger states required. Hence the proof.
\end{proof}

\begin{figure}
    \centering
    \includegraphics[width=0.9\linewidth]{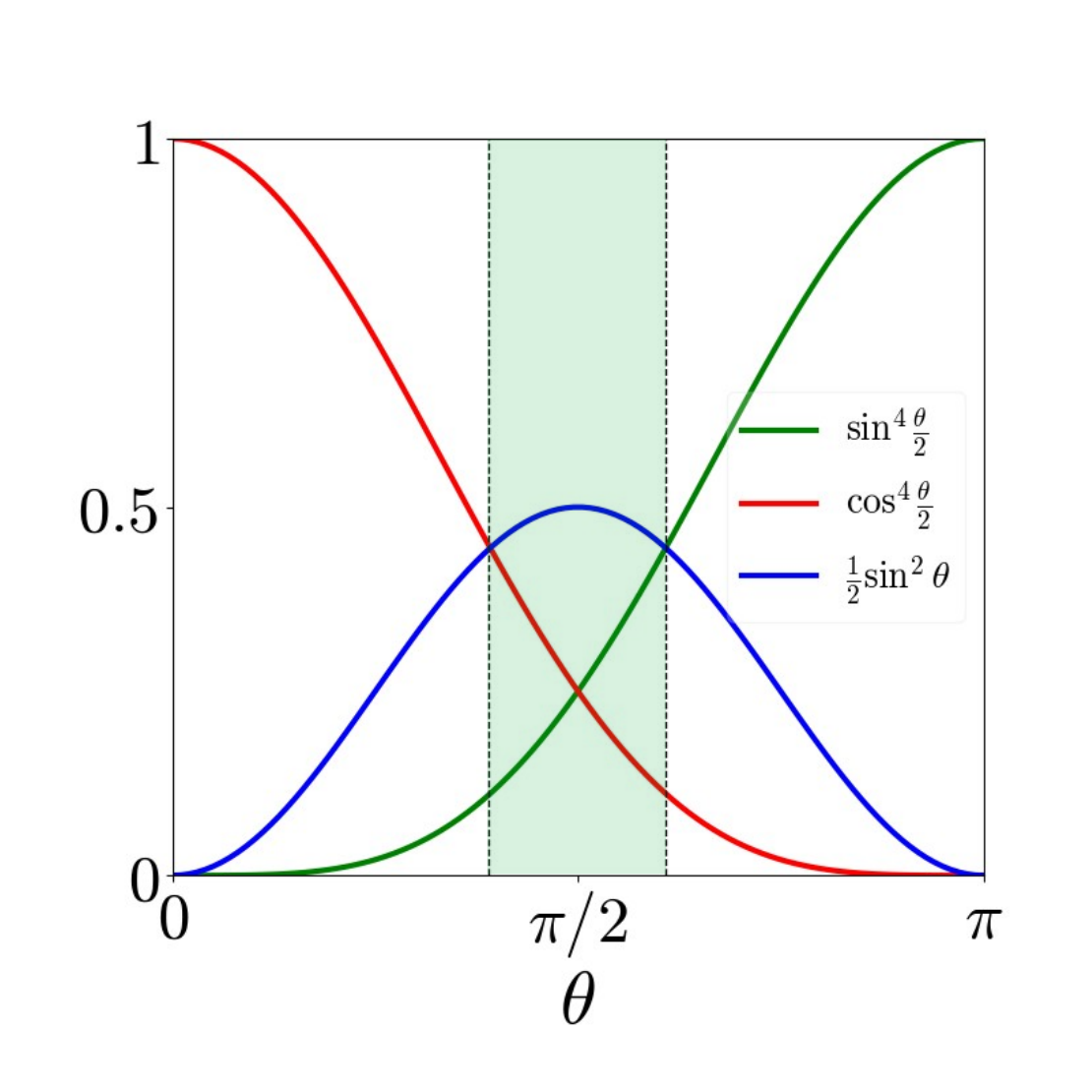}
    \caption{\textbf{Simultaneous charging.} Both the modes $b_1$ and $b_2$ charge within a finite range of $\theta$. See Proposition VIII. }
    \label{fig:theta_simultaneous}
\end{figure}

\begin{figure*}
    \centering\includegraphics[width=\linewidth]{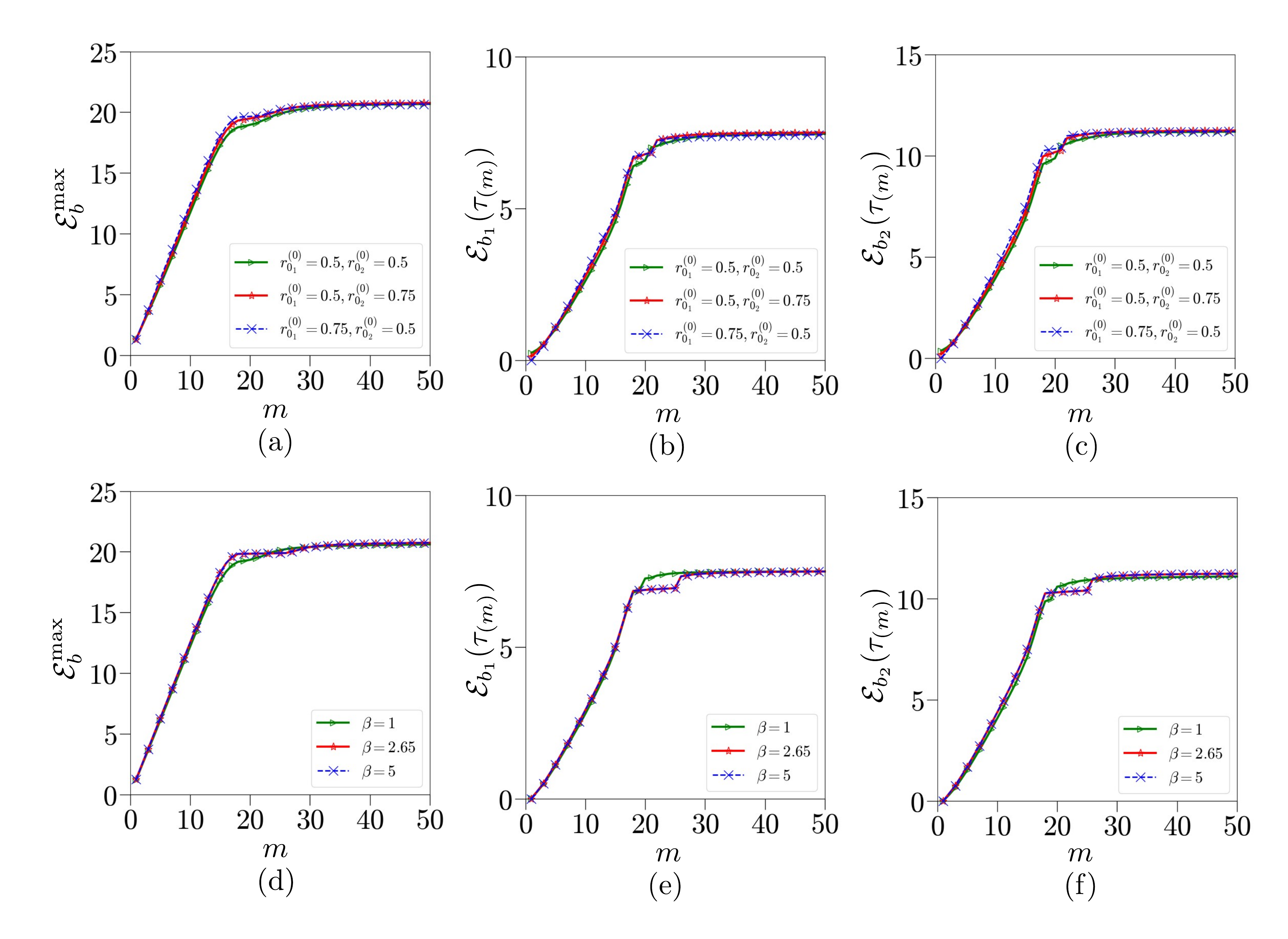}
    \caption{Variations of $\mathcal{E}^{\max}_b$, $\mathcal{E}^{\max}_{b_1}$, and $\mathcal{E}^{\max}_{b_2}$ with increasing $m$, starting from (a)-(c) truncated mixtures of two lowest lying states, and (d)-(f) thermal states. In each round, the duration of the time evolution is $\tau_{(m)}$, the time taken to obtain $\mathcal{E}^{\max}$ for the collective battery $b$.  }
    \label{fig:doublemode_multiple_rounds}
\end{figure*}

\noindent The finite range of $\theta$ corresponding to a set of charger initial states $\ket{\psi}$ is demonstrated in Fig.~\ref{fig:theta_simultaneous}. One can  continue repeating the charging cycle to charge both $b_1$ and $b_2$ simultaneously, following Sec.~\ref{subsec:repeated_charging}.

\emph{Does similar features occur for initial battery states other than the ground states?} To check this, we consider two specific cases: (i) In one,  both modes $b_1$ and $b_2$ are in states having the form  given in Eq.~(\ref{eq:truncated_states}) with $d=2$, and (ii) in the other, both $b_1$ and $b_2$ start from their respective canonical thermal states with equal temperature $\beta^{-1}$. In Fig.~\ref{fig:doublemode_multiple_rounds}(a)-(c), we plot $\mathcal{E}^{\max}_{b}$, $\mathcal{E}_{b_1}$, and $\mathcal{E}_{b_2}$ corresponding to the case (i) in each round of charging for different initial states $\rho_{b_i,2}^{\text{tr}}$ $(i=1,2)$.  It is clear from the figures that while the collective battery $b$ is charged in the first round, neither $b_1$ nor $b_2$ is charged in the same. However, repeating the charging rounds charge both modes simultaneously. The charging increases with $m$ at first, eventually reaching a saturation. Similar features are observed with thermal states as the initial states for $b_1$ and $b_2$ (see Fig.~\ref{fig:doublemode_multiple_rounds}(d)-(f)).

\section{Conclusion}
\label{sec:conclusion}

In this paper, we explored a bipartite setup of a battery and a charger, which is first evolved by a joint unitary operator generated by a Hamiltonian describing the battery-charger duo, followed by extracting energy from the battery by tracing out the charger. We showed that the minimum daemonic advantage that is achievable by alternatively performing a measurement on the charger, quantified by the daemonic ergotropy, can also be reached when the charger is discarded via tracing out. We define the difference between the minimum daemonic ergotropy and the maximum ergotropy obtained after tracing out the charger at the time when the ergotropy is maximum as the daemonic gap. We showed that a battery in the form of a harmonic mode, and a qubit charger interacting with the battery via the Jaynes-Cummings interaction can be designed, in which the daemonic gap vanishes with proper choice of the initial state of the battery, including the ground state, truncated mixture of low-lying levels, and canonical thermal states. Further, defining the charging cycle via tracing out the charger allows one to repeatedly apply the charging cycle. We showed that such repetition can be exploited to completely charge the battery along with accessing the full daemonic band, defined as  the difference between the maximum and the minimum daemonic ergotropy. We generalized the battery-charger design to the situation where the battery is constituted of multiple harmonic modes, and the charger is replaced by a qudit. We demonstrated that the repetition of the charging cycle is the key to simultaneously charge the component harmonic modes together while charging the battery collectively.  

Our work opens up a number of interesting questions. It is important to investigate whether gapless charging cycles can be universally designed for all possible initial states using specific models for the battery and charger. Further, instead of the resonator mode, the $d_b$-level system can be considered to be constituted of a number of interacting systems with each having a smaller Hilbert space dimension, eg. an interacting quantum spin model. While the complexity of the battery increases due to the involvement of the Hamiltonian parameters, it would be interesting to see if such charging cycles can be designed for these batteries as well.  Also, modifications of the performance of the charging cycle due to presence of imperfections in the battery state as well as in the charger measurement need to be explored.

\acknowledgements

The Authors acknowledge the use of \href{https://github.com/titaschanda/QIClib}{QIClib} -- a modern C++ library for general purpose quantum information processing and quantum computing. A.K.P acknowledges the support from the Anusandhan National Research Foundation (ANRF) of the Department of Science and Technology (DST), India, through the Core Research Grant (CRG) (File No. CRG/2023/001217, Sanction Date 16 May 2024).

\bibliography{ref}

\end{document}